# Origin of the non-monotonic variance of $T_c$ in the 1111 iron based superconductors with isovalent doping

Hidetomo Usui[*], Katsuhiro Suzuki & Kazuhiko Kuroki

Department of Physics, Osaka University, 1-1 Machikaneyama-cho, Toyonaka, Osaka, 560-0043, Japan
[*]email : h_usui@presto.phys.sci.osaka-u.ac.jp

**Motivated by recent experimental investigations of the isovalent doping iron-based superconductors $LaFe(As_xP_{1-x})O_{1-y}F_y$ and $NdFe(As_xP_{1-x})O_{1-y}F_y$ we theoretically study the correlation between the local lattice structure, the Fermi surface, the spin fluctuation-mediated superconductivity, and the composition ratio. In the phosphides, the $d_{XZ}$ and $d_{YZ}$ orbitals barely hybridize around the $\Gamma$ point to give rise to two intersecting ellipse shape Fermi surfaces. As the arsenic content increases and the Fe-As-Fe bond angle is reduced, the hybridization increases, so that the two bands are mixed to result in concentric inner and outer Fermi surfaces, and the orbital character gradually changes to $d_{xz}$ and $d_{yz}$, where $x$-$y$ axes are rotated by 45 degrees from $X$-$Y$. This makes the orbital matching between the electron and hole Fermi surfaces better and enhances the spin fluctuation within the $d_{xz/yz}$ orbitals. On the other hand, the hybridization splits the two bands, resulting in a more dispersive inner band. Hence, there is a trade-off between the density of states and the orbital matching, thereby locally maximizing the $d_{xz/yz}$ spin fluctuation and superconductivity in the intermediate regime of As/P ratio. The consistency with the experiment strongly indicate the importance of the spin fluctuation played in this series of superconductors.**

The discovery of the iron based superconductors[1,2] has given great impact to the field of condensed matter physics not only because of the high transition temperature ($T_c$) of superconductivity [3] but also because a lot of materials have been discovered with various blocking layers [4-8]. One of the most attractive features of the iron based superconductors is the correlation between the lattice structure and superconductivity. It has been found that the $T_c$ correlates with the Fe-$Pn$-Fe ($Pn$ = pnictogen) bond angle, and reaches its maximum at around 109 degrees by Lee *et al*. [9]. We have theoretically explained this correlation from the band structure point of view within the spin-fluctuation-mediated superconductivity [10,11]. There are basically three hole Fermi surfaces around (0,0) and ($\pi$, $\pi$), and two electron Fermi surfaces around ($\pi$,0) and (0, π) in the iron based superconductors where the bond angle is around 110 degrees. In this Fermi surface configuration, the spin susceptibility is enhanced because of the nesting between the hole and the electron Fermi surfaces, and superconductivity is optimized around 110 deg. corresponding to NdFeAsO. Upon reducing the bond angle from 110 deg., the hole Fermi surface around (π,π) shrinks and is eventually lost around 114 deg. (corresponding to LaFeAsO), so that the spin susceptibility and the superconductivity are suppressed as the bond angle is increased. The expectation from this view was that superconductivity is monotonically degraded with the increase of the bond angle.

Recently, 1111 iron based superconductors with isovalent doping have been experimentally investigated in $Ln$FeAs$_x$P$_{1-x}$O$_{1-y}$F$_y$ ($Ln$=La, Nd, Pr) [12-20]. The main effect of substituting phosphorus by arsenic is to vary the Fe-$Pn$-Fe bond angle while maintaining the electron content. Surprisingly and interestingly, it was found that $T_c$ takes its local maximum in the intermediate regime of arsenic/phosphorous ratio, which indicates that the superconductivity is locally optimized at a certain bond angle larger than 109 deg., i.e., between 113 deg. (arsenic end) and 120 deg. (phosphorus end), in contradiction with Lee's plot [9] as well as the previous theoretical expectations. Furthermore, when the electron doping is changed by controlling the fluorine content, the arsenic content at which the superconductivity is locally optimized moves to a higher concentration. Also, around the arsenic content where $T_c$ is locally optimized, strong spin fluctuation is observed [17-19]. The number of the Fermi surfaces does not change in the intermediate regime of the arsenic content, so that the previous study [10,11] cannot explain this local optimization of $T_c$. Therefore an unrevealed feature of the iron-based materials regarding this local optimization of superconductivity should be present.

Given this background, we study the correlation between the local lattice structure, the orbital character of the Fermi surface, and $T_c$ in the 1111 system with isovalent doping. In order to extract the essence of the isovalent doping, we calculate the band structure and study the superconductivity of the hypothetical lattice structure of LaFeAsO varying only the Fe-$Pn$-Fe bond angle. To our surprise, it is found that superconductivity is indeed locally optimized in the bond angle regime between 113 deg. and 120 deg., which corresponds to the intermediate regime of the arsenic (or phosphorous) content. The origin of this local optimization is traced back to the variation of the orbital character and the density of states of the hole Fermi surfaces around the Γ point, which is controlled by the bond angle. The consistency with the experiment strongly indicates the importance of the spin fluctuation played in this series of superconductors.

**Results**

**The correlation between the lattice structure and the band structure.** It is known that the Fermi surface of the iron based superconductors is mainly governed by the Fe-*Pn*-Fe bond angle, or the pnictogen height measured from the Fe plane [10,11]. To show this, we compare in Fig.1 the Fermi surface of NdFeAsO, NdFePO, and LaFePO calculated for experimental lattice structure [21] to those calculated for the hypothetical lattice structure of LaFeAsO with the Fe-*Pn*-Fe bond angle changed to the corresponding values, namely, 111, 118 and 120 degrees, respectively. It is seen that the essential features of the Fermi surface are nicely reproduced by the hypothetical lattice structure of LaFeAsO by varying only the bond angle. As studied previously [10,11], one of the effects of varying the bond angle appears in the number of Fermi surfaces ; the Fermi surface around the wave vector ($\pi,\pi$) is lost for large enough bond angles. It is known that the hole Fermi surface around ($\pi,\pi$) plays a key role in optimizing the superconductivity in the iron based superconductors.

In the present study, we put more focus on the difference of the Fermi surface between LaFeAsO (113 deg.) and LaFePO (120 deg.). The shape of the two hole Fermi surfaces around (0,0) is two concentric round ones in LaFeAsO, but two intersecting ellipse in LaFePO. This is because the major axis of the hole Fermi surfaces in LaFePO is *X* and *Y*, namely, the $d_{XZ}$ and $d_{YZ}$ orbitals are barely hybridized. On the other hand, in LaFeAsO the $d_{XZ}$ and $d_{YZ}$ orbitals are strongly hybridized and gives rise to the $d_{xz}$ and $d_{yz}$ orbital character on the two hole Fermi surfaces as shown in Fig.1(b). Here *X-Y* and *x-y* axes are rotated by 45 degrees, and the latter is the major axes of the unfolded Brillouin zone. The mixture of the $d_{xz/yz}$ is controlled by the bond angle, so that the relationship of the two Fermi surfaces between NdFeAsO and NdFePO is the same as explained above.

**Superconductivity.** Let us move on to the result of superconductivity. In order to extract the essence of the doping dependence, we adopt the hypothetical lattice structure of LaFeAsO varying only the bond angle. We show in Fig.2 the calculation result of the eigenvalue of the Eliashberg equation obtained by applying the Fluctuation Exchange approximation (FLEX) [22,23] to the models. The band fillings *n* = 6.125, 6.135, and 6.15 taken here corresponds to the electron doping ratio of 0.125, 0.135, and 0.15, respectively. The eigenvalue $\lambda_E$ of the Eliashberg equation reaches unity at the superconducting transition temperature $T = T_c$, so that we can use the eigenvalue as a qualitative measure for $T_c$. The result shows that superconductivity is locally maximized in the intermediate regime of the bond angle between 113 deg. (LaFeAsO) and 120 deg. (LaFePO), namely, the arsenic content. These results are in qualitative agreement with previous experimental results [16-18,20]. More interestingly, the bond angle at which the eigenvalue is maximized moves to smaller values when the electron doping level increases. This again is in qualitative agreement with the experimental results showing that the As content at which the $T_c$ is optimized moves toward larger values as the electron doping (F content) increases. Note that the band filling taken here and the fluorine doping rate in the experiment [20] do not correspond quantitatively, but the experimental materials may have oxygen deficiencies, so that the actual amount of the doped electrons may be larger than that expected from the fluorine doping rate $y$. We also show the intraorbital spin susceptibility of the $d_{xy}$ and $d_{xz/yz}$ orbitals for several bond angles in Fig. 3. The $d_{xy}$ spin susceptibility is monotonically suppressed as the bond angle increases (decreasing the arsenic doping rate $x$) because the hole Fermi surface around $(\pi,\pi)$ originating from the $d_{xy}$ orbital sinks below the Fermi level in the large bond angle regime. On the other hand, the $d_{xz/yz}$ spin susceptibility is locally maximized in the intermediate bond angle regime where superconductivity is locally maximized. This clearly shows that the $d_{xz/yz}$ orbitals are responsible for the local enhancement of the superconductivity. We also show the superconducting gap function of the electron Fermi surface in Fig. 3. The gap symmetry is nodal $s\pm$-wave because the Fermi surface around $(\pi,\pi)$ is absent in this bond angle regime, as discussed in ref.[10,24-31].

Now we discuss why the spin susceptibility of the $d_{xz/yz}$ orbitals is locally maximized in the intermediate bond angle regime. As described in the discussion of the band structure, one of the difference between arsenides (small bond angle) and phosphides (large bond angle) is the orbital component of the two hole Fermi surfaces around the wave vector $(0,0)$. For the phosphide, these hole Fermi surfaces consist of nearly unhybridized the $d_{XZ}$ and $d_{YZ}$ orbitals, while for the arsenide, they are strongly hybridized, so that they have $d_{xz}$ and $d_{yz}$ character depending on the portion of the Fermi surface as shown in Fig.1(b). On the other hand, the two electron Fermi surfaces around $(\pi,0)$ and $(0,\pi)$ always have the $d_{xz}$ and $d_{yz}$ orbital character as in Fig. 1(b). Therefore, in the large bond angle regime around LaFePO, the orbital character between hole and electron Fermi surfaces is different, so that the spin susceptibility and superconductivity are suppressed because of the "bad" nesting in the sense that the orbital character is not matched between electron and hole Fermi surfaces (Fig. 4(a)). Upon increasing the arsenic doping concentration, or decreasing the bond angle, the $d_{XZ}$-$d_{YZ}$ hybridization becomes stronger, and the two hole Fermi surfaces around $(0,0)$ tend to be constructed from the $d_{xz}$ and $d_{yz}$ orbital components. The good nesting (i.e. with matched orbital character) between the inner hole Fermi surface around (0,0) and the electron Fermi surfaces gives rise to strong spin fluctuations (Fig.4(a)), so that the spin susceptibility and superconductivity are both enhanced. As the arsenic content is further increased, the effective mass of the inner hole Fermi surface decreases. The comparison of the band structure between 110 deg. and 120 deg. in Fig. 4(b) shows that the inner band around (0,0) becomes dispersive as the As content increases (the bond angle becomes smaller) because of the larger splitting between the two bands due to the increased hybridization. Hence, there is a trade-off between the orbital matching of the nesting and the density of states of the inner Fermi surface. Therefore, the $d_{xz/yz}$ spin fluctuation and superconductivity is locally maximized around the intermediate bond angle regime, namely, intermediate arsenic content regime.

The electron doping (i.e., fluorine doping) dependence of the second dome can also be understood as follows. The electron doping dependence of the Fermi surface is shown in Fig. 4(c) ($n$ = 6 and 6.2). Upon increasing the electron doping, the hole Fermi surface becomes small. Moreover, the shape of the Fermi surface slightly changes from the round shape originating from the $d_{xz/yz}$ orbitals to the ellipse shape originating from the $d_{XZ/YZ}$. This is because the $d_{XZ}$-$d_{YZ}$ hybridization becomes weaker upon approaching the $\Gamma$ point, so that the $d_{XZ}/d_{YZ}$ character increases as the Fermi level is increased and the hole Fermi surfaces shrink. Therefore, the bond angle at which the superconductivity is locally maximized moves toward smaller values, namely, requires larger As content so as to increase the $d_{XZ}$-$d_{YZ}$ hybridization. In total, the experimental results of both the isovalent and the electron doping dependence of $T_c$ can be well explained within the nature of the hole Fermi surfaces around (0,0).

**Conclusion**

Motivated by recent experiments on $Ln$FeAs$_x$P$_{1-x}$O$_{1-y}$F$_y$ which observe non-monotonic evolution of the superconducting $T_c$ against the arsenic/phosphorous ratio, we have studied the correlation between the local lattice structure, the orbital character of the Fermi surface, and superconductivity. It is found that one of the key points is the orbital component of the two hole Fermi surfaces around the wave vector (0,0). The orbital character of these hole Fermi surfaces gradually changes from the $d_{XZ/YZ}$ to the $d_{xz/yz}$ orbitals because of the hybridization between the $d_{XZ}$ and $d_{YZ}$ orbitals is enhanced as the arsenic content increases and the bond angle is reduced. The electron Fermi surfaces are constructed from the $d_{xz/yz}$ orbitals, so that the orbital character matching between electron and hole Fermi surfaces gets better as the arsenic content is increased, and this itself enhances superconductivity. On the other hand, the increase of the $d_{XZ}$-$d_{YZ}$ hybridization reduces the density of states of the inner hole Fermi surface, so that there is a trade-off between the two effects. Hence, the spin susceptibility and superconductivity are locally optimized in the intermediate arsenic content regime. As the amount of the electron doping increases, the bond angle at which $T_c$ is locally maximized decreases, which is also consistent with the experiment.

The agreement with the experiments strongly indicates that the nesting condition of the portion of the Fermi surfaces originating from the $d_{xz/yz}$ orbitals plays a key role in the $T_c$ evolution of the isovalent doped 1111 system. On the other hand, recently we have also found [32] that the $d_{xy}$ orbital plays an important role in the strong enhancement of the $T_c$ in hydrogen doped 1111 system [33]. There, it has been shown that good Fermi surface nesting is not important, and the peculiar relation between the nearest and next nearest neighbor electron hoppings in real space is the key [34-37]. From the previous and present studies, we can schematically depict in Fig.5 the entire "phase diagram" of the 1111 system regarding the spin fluctuation/ordering and superconductivity. The two strongly different nature of the pairing glue found in a single family, both originating from the spin fluctuations, is a hallmark of a multiorbital superconductor. What makes the iron-based special among the multiorbital superconductors is that the pairing glue originating from different orbitals favor the same pairing ($s\pm$) state. This peculiar feature of the iron-based superconductors is likely to be one of the main reasons for the high $T_c$.

**Method**

**First principles band calculation.** We calculate the band structure using the VASP package [38]. The Projector Augmented Wave method (PAW) [39] is used to calculate the wave functions, and the Generalized Gradient Approximation (GGA) formulated by Perdew, Burke and Ernzerhof (GGA-PBE [40]) is taken into account for the exchange-correlation functional. Virtual crystal approximation is used to take into account the effect of partial substitution of phosphorus by arsenic. In the first-principles calculation, we take 10×10×10 k-points and 600eV cut off energy. After the first principles band calculation, we construct Maximally localized Wannier functions [41] using wannier90 package [42] through the VASP package. In wannier90, we take 8×8×8 k-points.

**Superconductivity.** We consider the standard muti-orbital interactions, namely, the intraorbital $U$, the inter-orbital $U'$, the Hund's coupling $J$, and the pair hopping interaction $J'$. In the fluctuation exchange (FLEX) approximation [22,23], we take into account bubble and ladder type diagrams consisting of renormalized Green's functions to obtain the susceptibilities, which are used to calculate the self-energy. We determine the renormalized Green's functions from the Dyson's equation. The obtained Green's function is plugged into the linearized Eliashberg equation, whose eigenvalue $\lambda_E$ reaches unity at the superconducting transition temperature $T = T_c$. Since the three dimensionality is not strong in LaFeAsO, we take a two-dimensional model where we neglect the out-of-plane hopping integrals, and take 32×32 k-point meshes and 8192 Matsubara frequencies at temperature $T = 0.005$eV.

As for the electron-electron interaction values, we adopt the orbital-dependent interactions as obtained from first-principles calculation in Ref. [43] of LaFeAsO. As has been studied in Refs. [44-46], the FLEX for models obtained from LDA calculations tend to overestimate the effect of the self-energy because LDA already partially takes into account the effect of the self-energy in the exchange-correlation functional. Therefore we subtract the self-energy at lowest Matsubara frequency from FLEX calculations in each self-consistent loop as in ref. [32]. We also intend to reduce the double counting of the electron correlation effect by enlarging by 50% the bare band width of the model Hamiltonian compared to that obtained from the first principles calculation (multiplying all the hoppings by 1.5).

**Acknowledgements**

We are grateful to H. Mukuda, S. Tajima, S. Miyasaka, A. Takemori, K. T. Lai, K. Tanaka, H. Arai, and Y. Fuseya for many fruitful discussions and A. Takemori, S. Miyasaka and S. Tajima for providing the experimental lattice structure parameter. Numerical calculations were performed at the facilities of the Supercomputer Center, Institute for Solid State Physics, University of Tokyo. This study has been supported by Grants-in-Aid for Scientific Research No. 24340079 and No. 25009605 from the Japan Society for the Promotion of Science.

**Author contribution statement**

H.U. performed the first principles band calculation and constructed 5 orbital models. H.U. and K.K. calculated superconductivity. K.S. constructed the FLEX calculation code. H.U, K.K and K.S. discussed the results. H.U. and K.K. wrote the manuscript.

**Additional information**

Competing financial interests: The authors declare no competing financial interests

**Figure legends**

**Figure 1** (a) The Fermi surface of NdFeAsO, LaFePO and NdFePO. The Fermi surfaces of the hypothetical lattice structure of LaFeAsO are superposed by dotted lines with the Fe-*Pn*-Fe bond angle changed to the corresponding values, 111, 118 and 120 degrees, respectively. (b) The orbital component of the Fermi surface of the hypothetical lattice structure of LaFeAsO with the bond angle of 120 deg. (left) and 113 deg. (right).

**Figure 2** The eigenvalue $\lambda_E$ of the Eliashberg equation for the hypothetical lattice structure of LaFeAsO with varying the Fe-*Pn*-Fe bond angle. Three values are taken for the electron concentration *n*, where *n*-6 is the electron doping ratio.

**Figure 3** The spin susceptibility of the $d_{xy}$ and $d_{xz/yz}$ orbitals (left panels) and the superconducting gap of the electron Fermi surfaces (right) at band filling $n = 6.135$.

**Figure 4**   (a) the nesting vector and the orbital component of the Fermi surface of the hypothetical lattice structure of LaFeAsO for the Fe-*Pn*-Fe bond angle of 120 deg. (left) and 113 deg. (right). Schematic figures of the band structure and the Fermi surface with (b) isovalent doping and (c) electron doping. The band structure of hypothetical lattice structure of LaFeAsO is shown in the right panel of (b).

**Figure 5** A schematic phase diagram of 1111 regarding spin fluctuation/ordering and superconductivity, including both the electron and the isovalent doping systems.

**Figure 1**

(a)

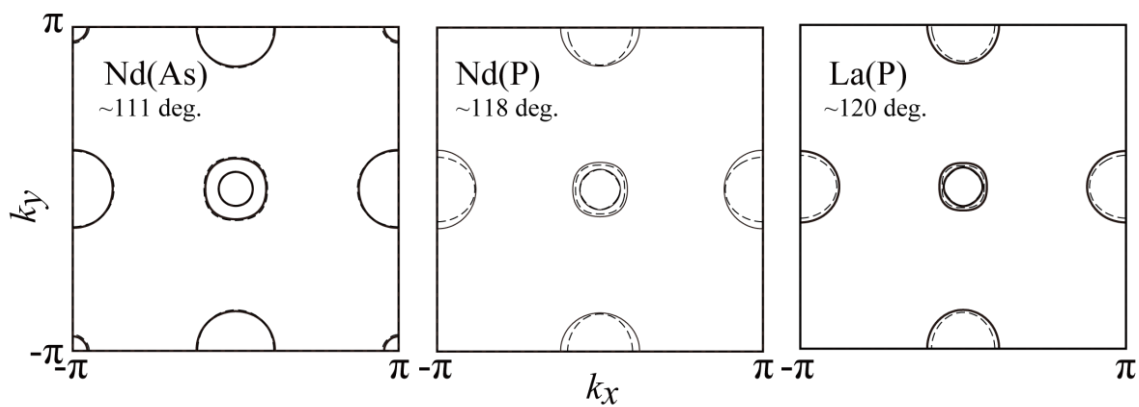

(b)

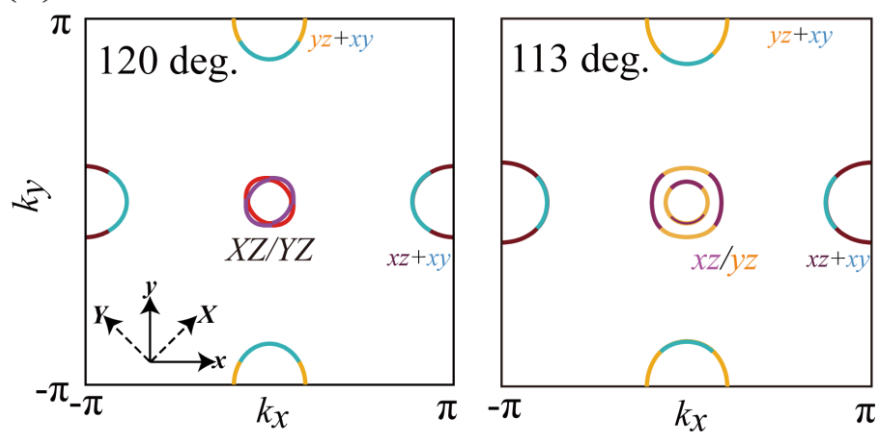

**Figure 2**

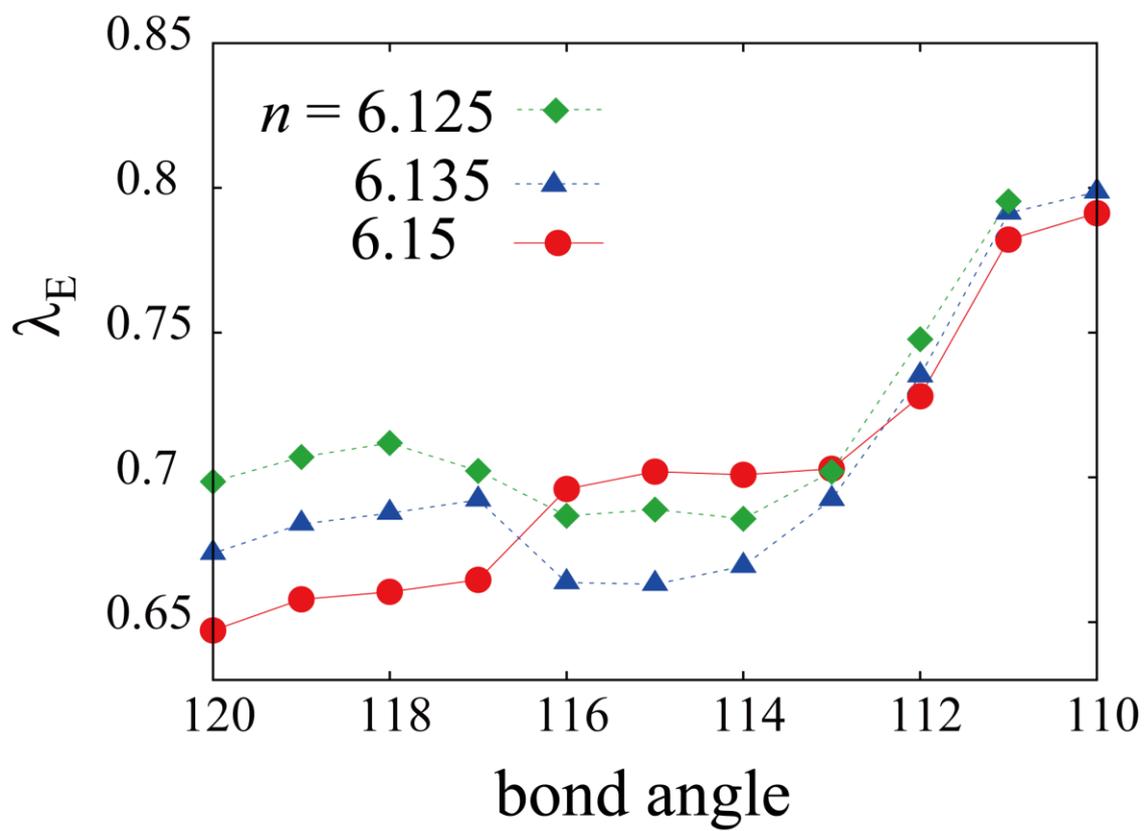

**Figure 3**

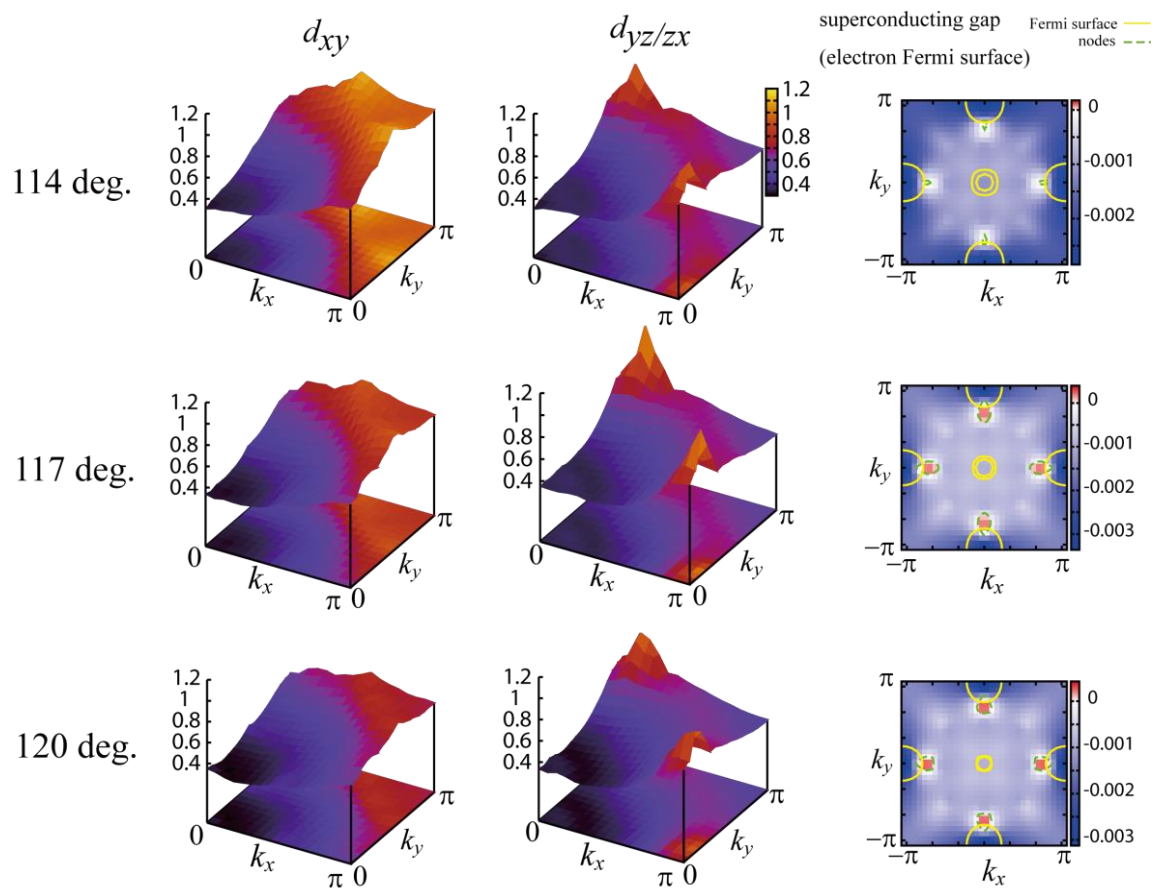

**Figure 4**

(a)

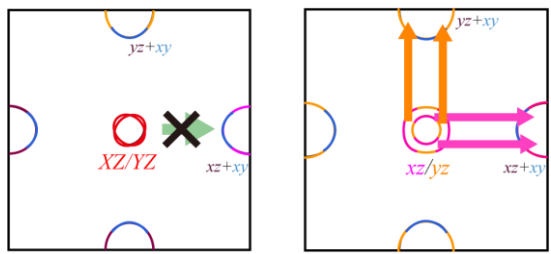

(b) isovalent doping

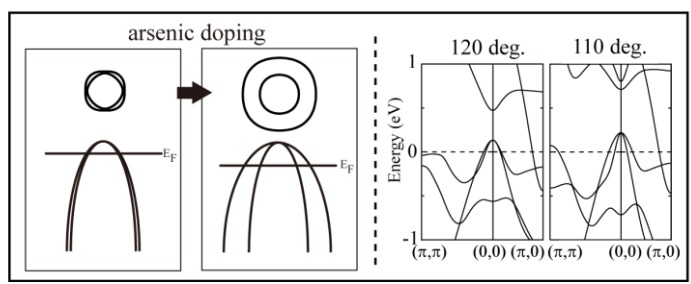

(c) electron doping

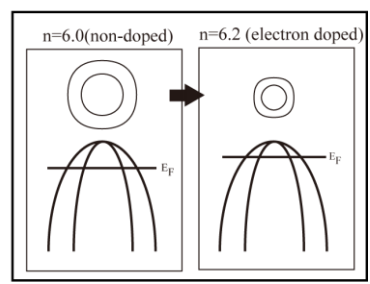

**Figure 5**

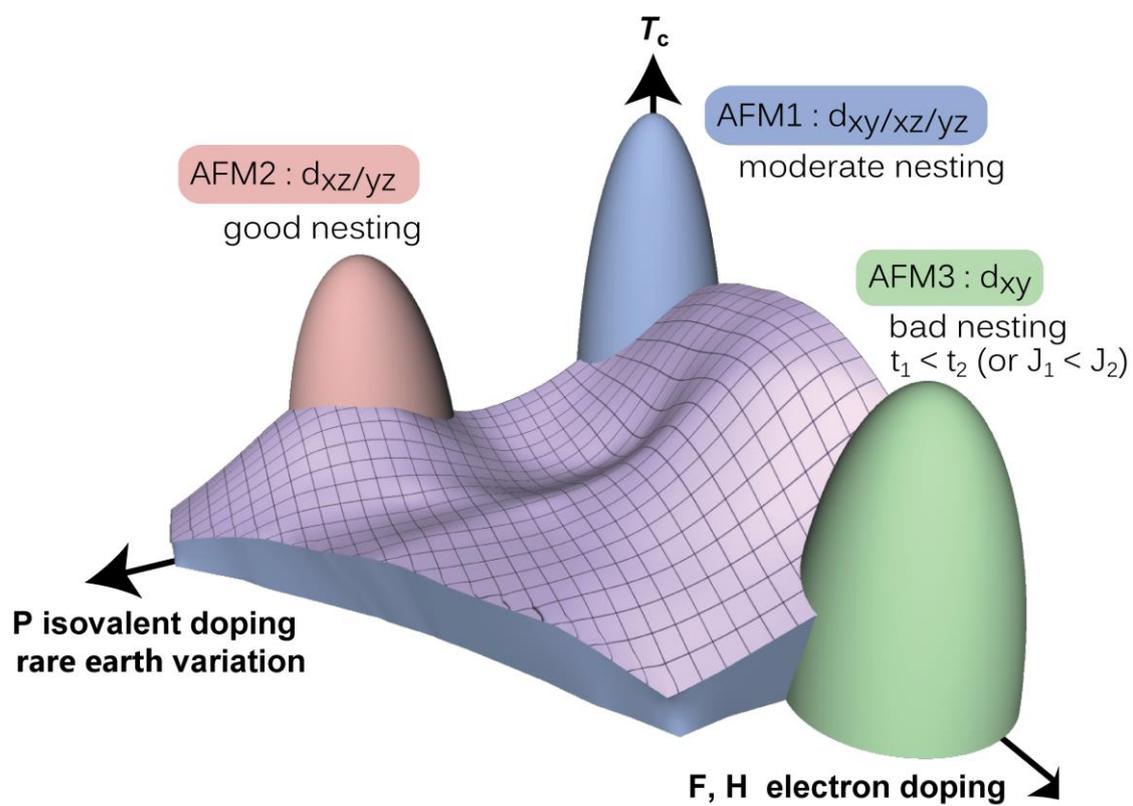